\documentclass[journal=jacsat,manuscript=article]{achemso}
\SectionNumbersOn
\usepackage[version=3]{mhchem} %
\usepackage{ifthen}
\usepackage{braket}
\usepackage{amsmath}
\usepackage{multirow}
\usepackage{makecell}
\usepackage{adjustbox}
\usepackage{subfigure}
\usepackage[title]{appendix}
\usepackage{tikz}

\newcommand{\blue}[1]{{#1}}
\newcommand{\red}[1]{{#1}}
\newcommand{\hl}[1]{{#1}}

\usepackage{hyperref}
\hypersetup{colorlinks=true,linkcolor=blue,citecolor=blue,urlcolor=blue}

\newcommand{\Refx}[1]{Ref.~\citenum{#1}}

\newcommand*{\prn}[1]{\left({#1}\right)}
\newcommand*{\sqbr}[1]{\left[{#1}\right]}
\newcommand*{\deriv}[3]{
    \ifthenelse{\equal{#1}{1}}
        {\frac{\mathrm{d}{#2}}{\mathrm{d}{#3}}}
        {\frac{\mathrm{d}^{#1}{#2}}{\mathrm{d}{#3}^{#1}}}}
\newcommand*{\pderiv}[3]{
    \ifthenelse{\equal{#1}{1}}
        {\frac{\partial{#2}}{\partial{#3}}}
        {\frac{\partial^{#1}{#2}}{\partial{#3}^{#1}}}
}

\newcommand*{\conc}[1]{\sqbr{\mathrm{#1}}}
\newcommand*{\ex}[1]{\mathrm{e}^{#1}}

\author{Ziyan Ye}
\affiliation[FDU]
{Department of Chemistry, Shanghai Key Laboratory of Molecular Catalysis and Innovative Materials, State Key Laboratory of Porous Materials for Separation and Conversion, Fudan University, Shanghai 200438, P. R. China}
\author{Eric R. Heller}
\affiliation
{Department of Chemistry, University of California, Berkeley, 94720 Berkeley, USA}
\email{hellere@berkeley.edu}
\author{Dong H. Zhang}
\affiliation[DICP]
{State Key Laboratory of Molecular Reaction Dynamics, Dalian Institute of Chemical Physics, Chinese Academy of Sciences, Dalian 116023, P. R. China}
\alsoaffiliation{%
University of Chinese Academy of Sciences, 
Beijing 100049, China
}%
\author{Jeremy O. Richardson}
\affiliation[ETH]
{Department of Chemistry and Applied Biosciences, ETH Z\"urich, Z\"urich 8093, Switzerland}
\email{jeremy.richardson@phys.chem.ethz.ch}
\author{Wei Fang}
\affiliation[FDU]
{Department of Chemistry, Shanghai Key Laboratory of Molecular Catalysis and Innovative Materials, State Key Laboratory of Porous Materials for Separation and Conversion, Fudan University, Shanghai 200438, P. R. China}
\email{wei_fang@fudan.edu.cn}

\title[Instantons at near-barrier crossings]
  {
  Instanton Theory for Nonadiabatic Tunneling through Near-Barrier Crossings
  }

\abbreviations{}

\begin{document}

\begin{abstract}
Many reactions in chemistry and biology involve multiple electronic states, rendering them nonadiabatic in nature. 
These reactions can be formally described using Fermi's golden rule (FGR) in the weak-coupling limit.
Nonadiabatic instanton theory presents a semiclassical approximation to FGR, which is directly applicable to molecular systems. 
However, there are cases where the theory has not yet been formulated.
For instance, in many real-world reactions including spin-crossover or proton-coupled electron transfer, the nonadiabatic crossing occurs near a potential energy barrier.
This scenario gives rise to competing nonadiabatic reaction pathways, some of which involve tunneling through a round-top barrier while simultaneously switching electronic states.
To date, no rate theory is available for describing tunneling via these unconventional pathways.  
Here we extend instanton theory to model this class of processes, which we term the ``non-convex'' regime. 
Benchmark tests on model systems show that the rates predicted by instanton theory are in excellent agreement with quantum-mechanical FGR calculations.
Furthermore, the method offers new insights into multi-step tunneling reactions and the competition between sequential and concerted nonadiabatic tunneling pathways.
\end{abstract}

\section{Introduction}

Nonadiabatic effects arise when multiple electronic states are involved in chemical reactions or other physical processes, leading to the breakdown of the Born--Oppenheimer approximation \cite{gonzalez_quantum_2021}.
Such effects are widespread in molecular science, with notable examples including spin-crossover reactions in transition-metal chemistry \cite{harvey_spin-forbidden_2014},
charge transfer in biochemistry \cite{marcus_electron_1985,brookes_quantum_2017} and dynamics through conical intersections in photochemistry \cite{yarkony_nonadiabatic_2012}.
Simulating nonadiabatic phenomena is critical for understanding important chemical, physical, and biological processes at the molecular level. 
However, atomistic simulations beyond the Born--Oppenheimer approximation pose significant challenges \cite{butler_chemical_1998}.

Fermi's golden rule (FGR) \cite{zwanzig_nonequilibrium_2001} gives the exact nonadiabatic quantum transition rate in the weak-coupling limit.
It has been widely applied to simulate processes like electron transfer, \cite{UlstrupBook} proton-coupled electron transfer (PCET) \cite{hammes-schiffer_theory_2010} and nonradiative transitions \cite{Jortner1971radiationless}. 
However, despite its reliability and broad applicability, FGR suffers from its tremendous computational cost in solving vibrational wave-functions, making its direct application to complex high-dimensional molecular systems unfeasible.
Instead of evaluating and summing over contributions from each possible pair of reactant and product vibrational states, it would be simpler and conceptually more enlightening to have a transition-state theory, in which the rate is approximated by a single term, which captures the dominant reaction mechanism.

The most widely used rate theory for nonadiabatic processes is the well-known Marcus theory \cite{marcus_theory_1956, marcus_exchange_1960}, which can be derived from FGR using classical statistical mechanics for the nuclei and assuming parabolic free-energy surfaces \cite{marcus_electron_1985}. 
Nonadiabatic transition state theory (NA-TST) \cite{lorquet_nonadiabatic_1988, cui_spin-forbidden_1999, harvey_spin-forbidden_1999, harvey_spin-forbidden_2014} goes beyond the assumption of globally parabolic surfaces by identifying the minimum-energy crossing point (MECP) and using the molecular partition functions evaluated at the MECP and the reactant minimum. 
Simple tunneling corrections to NA-TST have been proposed, such as the weak-coupling approximation (WC) \cite{lykhin_nonadiabatic_2016}, which assumes linear diabatic potential-energy surfaces (PESs) along the tunneling coordinate \cite{delos_reactions_1973}. While this approach is appropriate for the shallow-tunneling regime, it yields qualitatively wrong results in the low-temperature, deep-tunneling regime \cite{heller_spin_2021, heller_heavy-atom_2022}. 

Although semiclassical instanton theory (SCI) was originally formulated within the Born--Oppenheimer approximation\cite{Miller1975semiclassical},
it has been generalized to the nonadiabatic case,
\cite{Cao1997nonadiabatic,richardson_semiclassical_2015, richardson_ring-polymer_2015, heller_instanton_2020, ansari_instanton_2022, GRperspective}  where it provides a semiclassical approximation to FGR that can be directly applied to multidimensional molecular reactions.
The approach is based on Feynman's path-integral formulation of quantum mechanics \cite{feynman_quantum_1995} and estimates the transition rate based on a single optimal tunneling pathway, called an instanton, rather than requiring the evaluation of numerous overlaps between vibrational wavefunctions as in the original FGR\@.
The optimization of an instanton pathway is facilitated by the well-known ring-polymer discretization. \cite{richardson_ring-polymer_2018} 
In this formulation, the instanton path is located by searching for the saddle point of the ring polymer, and is thus a simple extension of a TST calculation. 
This makes instanton calculations far more computationally efficient than their quantum-mechanical counterparts, and it can
be combined with \textit{ab initio} electronic-structure theory, making it practical for large molecular systems without resorting to any approximate models.
Nonetheless, SCI predicts rates nearly identical to those from FGR, \cite{richardson_semiclassical_2015, heller_instanton_2020, ansari_instanton_2022} and has been successfully applied to several molecular systems, demonstrating excellent agreement with experimental results and providing evidence that nuclear tunneling may occur even for heavy atoms at room temperature \cite{heller_spin_2021, heller_heavy-atom_2022, fang_competing_2023, ansari_heavy-atom_2024, carbenes}.
Note that SCI requires diabatic states, which can be naturally defined by multiplicity in spin-crossover reactions or by constrained density functional theory \cite{wu_direct_2005} in charge transfer processes, or alternatively through various diabatization protocols. \cite{thiel_proposal_1999, shu_diabatization_2020, guan_high-fidelity_2021}

\begin{figure}[!b]
    \centering
    \subfigure{\includegraphics[width=3.3in]{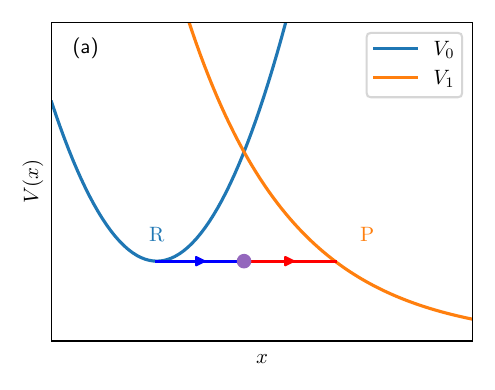}}
    \subfigure{\includegraphics[width=3.1in]{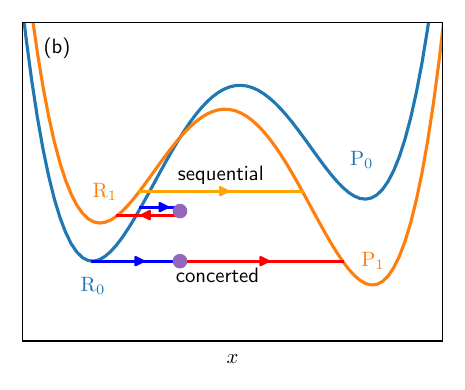}}
    \caption{
    Schematic diagrams of crossing diabatic states (0 and 1) in the convex (a) and non-convex (b) regimes along with the instantons involved. 
    (a) A system without any diabatic barrier. 
    The only possible reaction is the transition from reactants (R) to products (P). 
    The instanton consists of tunneling paths on either state (in blue and red) and changes electronic state at the hopping point (purple dot). 
    (b) A system with a near-barrier crossing. 
    Both potentials exhibit barriers, which lie in the vicinity of the crossing point. 
    Hence, there exist multiple competing instanton pathways. 
    }
    \label{fig:schematic}
\end{figure}
However, the complexity of real-world reactions often challenges the validity of existing nonadiabatic rate theories.
Reaction profiles involving the intersection of two double-well potentials, as illustrated in Fig.~\ref{fig:schematic}b, are commonly observed in PCET \cite{gagliardi_integrating_2010, zhu_proton-coupled_2024, hutchison_nonadiabatic_2024} and spin-crossover systems such as carbenes \cite{pharr_photochemistry_2023}, nitrenes \cite{besora_understanding_2008, nunes_heavy-atom_2020, li_water_2023}, diradicals \cite{viegas_spin-forbidden_2021} and transition-metal complexes \cite{harvey_spin-forbidden_2014, johansson_quantum_2007, usharani_theory_2013, canale_two_2016, zhou_sequential_2017, dergachev_predicting_2023}.
Despite the rich literature on nonadiabatic processes, to the best of our knowledge, reaction mechanisms (especially those that involve tunneling) with a near-barrier crossing remain largely unexplored. 
A major reason is that these systems are challenging for existing nonadiabatic rate theories. 
For example, NA-TST and WC only utilize local information in the neighborhood of the minimum-energy crossing point (MECP), without recognizing the complex shape of the reaction profile.
Although these theories generally suffice at high temperatures for simpler nonadiabatic reactions (as illustrated in Figure~\ref{fig:schematic}a), for the more complicated intersecting double-well system (Figure~\ref{fig:schematic}b), they can only describe the process from $\mathrm{R}_0$ to $\mathrm{R}_1$ (or conversely).
Therefore, it has long been believed that reactions occur via a sequential mechanism: first a state-changing transition from $\mathrm{R}_0$ to $\mathrm{R}_1$ followed by a state-preserving transition on the product surface to $\mathrm{P}_1$.
However, as we will show, a concerted deep-tunneling path may exist, directly linking $\mathrm{R}_0$ and $\mathrm{P}_1$ and rivaling the conventional sequential picture.

In this work, we apply instanton theory to systematically investigate tunneling mechanisms in nonadiabatic reactions in cases where the crossing occurs close to a barrier top.
Surprisingly, the standard nonadiabatic instanton rate expression is found to be not directly applicable to these systems due to an unexpected sign error in the prefactor, which describes fluctuations around the instanton path.
In the following, we extend SCI to this ``non-convex'' regime by analytic continuation. 
In fact, part of the theory we derive in this work was recently applied to explain the tunneling mechanisms of triplet carbenes, which demonstrates the practicalities of the method applied to realistic systems.\cite{carbenes}
Here, we present a rigorous justification of the approach, provide additional methodological details and introduce extensions to the inverted regime, along with a broader discussion of the scope of the non-convex case.

\section{Semiclassical instanton theory}
\label{sec:instanton}

In this section, we briefly summarize the previous derivation of SCI presented in Ref.~\citenum{ansari_instanton_2022}, which is valid in both the normal and inverted regimes. We consider a molecular system with reactant and product diabatic electronic states $\ket{0}$ and $\ket{1}$, whose Hamiltonian is given by
\begin{equation}
    \hat{H}=\hat{H}_0\ket{0}\bra{0}+\hat{H}_1\ket{1}\bra{1}+\hat{\Delta}\left(\ket{0}\bra{1}+\ket{1}\bra{0}\right),
\end{equation}
where $\hat{\Delta}=\Delta(\hat{x})$ denotes the diabatic coupling between the two states. The nuclear Hamiltonians are given by 
\begin{equation}
    \hat{H}_n=\frac{\hat{p}^2}{2m}+V_n\left(\hat{x}\right),
\end{equation}
where $n\in\left\{0,1\right\}$, and $\hat{x}$ comprises the $f$ nuclear degrees of freedom moving on PES $V_n$ according to their conjugate momenta $\hat{p}$. 
The variables are assumed to be mass-weighted such that all degrees of freedom have the same mass, $m$.
In the golden-rule regime, where the coupling $\hat{\Delta}$ is weak, the interaction can be treated perturbatively, leading to the well-known FGR
\begin{equation}
    kZ_0=\frac{2\pi}{\hbar}\sum_\mu\mathrm{e}^{-\beta E_0^\mu}\sum_\nu\left|\braket{\mu|\hat{\Delta}|\nu}\right|^2\delta\left(E_1^\nu-E_0^\mu\right), \label{equ:FGR}
\end{equation}
where $k$ is the reaction rate constant, $Z_0$ is the reactant partition function, $\beta=1/k_\text{B}T$ is the inverse temperature, $\mu$ and $\nu$ denote vibrational eigenstates on electronic states 0 and 1, and energy conservation is enforced by the $\delta$-function. 
Equation~\eqref{equ:FGR} can be equivalently expressed as an integral of the flux correlation function $c_\mathrm{ff}$ over time $t$, \cite{Miller1983rate, Wolynes1987nonadiabatic, ChandlerET}
\begin{equation}
    kZ_0=\int_{-\infty}^{+\infty}c_\mathrm{ff}\left(\tau+\mathrm{i}t\right)\mathrm{d}t, \label{equ:quantum rate}
\end{equation}
\begin{equation}
    c_\mathrm{ff}\left(\tau+\mathrm{i}t\right)=\frac{1}{\hbar^2}\mathrm{Tr}\left[\hat{\Delta}\,\mathrm{e}^{-\left(\beta\hbar-\tau-\mathrm{i}t\right)\hat{H}_0/\hbar}\hat{\Delta}^\dagger\,\mathrm{e}^{-\left(\tau+\mathrm{i}t\right)\hat{H}_1/\hbar}\right], \label{equ:corr func}
\end{equation}
where the integral in Eq.~\eqref{equ:quantum rate} is independent of the imaginary time $\tau$ by virtue of Cauchy's integral theorem, provided that the integration contour in the complex-time plane does not enclose any singularities. \cite{ansari_heavy-atom_2024}

SCI can now be derived as a semiclassical approximation to FGR. \cite{richardson_semiclassical_2015, heller_instanton_2020, ansari_instanton_2022}
The main idea is to replace the quantum propagators by their semiclassical counterparts, and then perform steepest-descent integration.
We expand the trace in Eq.~\eqref{equ:corr func} in a basis of position states and approximate the quantum propagators by the corresponding semiclassical van-Vleck propagators \cite{miller_classical_1971, gutzwiller_chaos_1990} to give
\begin{equation}
    kZ_0\sim\frac{1}{\hbar^2}\iiint \Delta\left(x'\right)\Delta^*\left(x''\right)\frac{\sqrt{C_0}\sqrt{C_1}}{\left(2\pi\hbar\right)^f}\,\mathrm{e}^{-S/\hbar}\,\mathrm{d}x'\mathrm{d}x''\mathrm{d}t, \label{equ:semiclassical rate}
\end{equation}
where the total path action $S\equiv S_0\left(x',x'',\beta\hbar-\tau-\mathrm{i}t\right)+S_1\left(x'',x',\tau+\mathrm{i}t\right)$ consists of an action integral along a classical trajectory in complex time $z_n$ on each electronic state
\begin{equation}
    S_n\left(x_\mathrm{i},x_\mathrm{f},z_n\right)= \int_0^{z_n}\left[\frac{1}{2}m\left|\Dot{x}\left(u\right)\right|^2+V_n\left(x\left(u\right)\right)\right]\mathrm{d}u,
\end{equation}
with $x_\mathrm{i}$ and $x_\mathrm{f}$ being the starting and ending points ($x_\mathrm{i}=x'',\ x_\mathrm{f}=x'$ for $n=0$, and $x_\mathrm{i}=x',\ x_\mathrm{f}=x''$ for $n=1$), and $C_n$ is defined by the $f\times f$ determinant
\begin{equation}
    C_n=\left|-\frac{\partial^2 S_n}{\partial x_\mathrm{i}\partial x_\mathrm{f}}\right|.
\end{equation}
The semiclassical instanton rate expression is then obtained by evaluating the integrals in Eq.~\eqref{equ:semiclassical rate} using the steepest-descent approximation around the instanton, a periodic trajectory satisfying the principle of stationary action.

\red{The method of steepest descent provides an asymptotic approximation of $f$-dimensional integrals of the form
\begin{align}
    \label{equ:sd}
    \notag \int A(v)\,\mathrm{e}^{-\phi(v)/\hbar}\,\mathrm{d} v&\sim\int A(v)\,\mathrm{e}^{-\phi(\widetilde{ v})/\hbar-\frac{1}{2\hbar}(v-\widetilde{ v})^\mathrm{T}\phi''(\widetilde{v})( v-\widetilde{v})}\,\mathrm{d} v \\
    &=\frac{(2\pi\hbar)^{f/2}}{\sqrt{\det\phi''(\widetilde{v})}}A(\widetilde{v})\,\mathrm{e}^{-\phi(\widetilde{v})/\hbar},
\end{align}
by locating the saddle point $\tilde{v}$ of the exponent, deforming the complex contour so that it passes through the saddle point along the direction of steepest descent, expanding the exponent to second order about this point, $\phi''\equiv\frac{\mathrm{d}^2\phi}{\mathrm{d}v\mathrm{d}v}$, and then evaluating the resulting Gaussian integral.\cite{BenderBook}
}

Carrying out the position and time integrals in Eq.~\eqref{equ:semiclassical rate} by steepest descent leads to the semiclassical instanton rate expression \cite{richardson_semiclassical_2015,heller_instanton_2020}
\begin{equation}
    k_\mathrm{SCI}Z_0=\sqrt{2\pi\hbar}\frac{\Delta^2}{\hbar^2}\frac{\sqrt{C_0}\sqrt{C_1}}{\sqrt{C}}\left(-\frac{\mathrm{d}^2 S}{\mathrm{d}\tau^2}\right)^{-\frac{1}{2}}\mathrm{e}^{-S/\hbar}, \label{equ:instanton}
\end{equation}
where all quantities are evaluated at the instanton, which is a stationary point of $S$ with respect to $x'$, $x''$ and $\tau$. 
The prefactor $C$ is given by
\begin{equation}
    C=
    \begin{vmatrix}
        \frac{\partial^2 S}{\partial x'\partial x'} & \frac{\partial^2 S}{\partial x'\partial x''} \\
        \frac{\partial^2 S}{\partial x''\partial x'} & \frac{\partial^2 S}{\partial x''\partial x''}
    \end{vmatrix}.
\end{equation}
The minus sign in front of $\frac{\mathrm{d}^2S}{\mathrm{d}\tau^2}$ arises from the Cauchy--Riemann equations, which relate the derivatives with respect to real time $t$ to those with respect to imaginary time $\tau$. 
\red{The diabatic coupling is defined as $\Delta\equiv\left|\Delta\left(x'\right)\right|=\left|\Delta\left(x''\right)\right|$. Here, we use the property $x'=x''$ of the instanton, which is a well-established result in previous Born--Oppenheimer\cite{richardson_ring-polymer_2018,Perspective,Rommel2011grids,methanol} and golden-rule instanton studies\cite{richardson_semiclassical_2015,richardson_ring-polymer_2015,MLJ,heller_spin_2021} in the absence of conical intersections\cite{fang_competing_2023}. This equivalence reflects a general feature of such theories, which are derived from the linear response of a system at thermodynamic equilibrium. As a consequence, the reaction rates obey detailed balance, and the optimal reaction path is identical for the forward and backward transitions between two metastable states.}\cite{NEQI}

Note that in Eq.~\eqref{equ:instanton}, we write the square roots separately to emphasize that, in this derivation, $\boldsymbol{C}_0$, $\boldsymbol{C}_1$ and $\boldsymbol{C}$ (the matrices corresponding to determinants $C_0$, $C_1$ and $C$) are each required to be positive definite, while $\frac{\mathrm{d}^2S}{\mathrm{d}\tau^2}$ must be negative, in accordance with the steepest-descent method. 
All of these conditions hold for systems in the normal regime when the curvatures of the diabatic surfaces are predominately positive near the crossing.
\cite{ansari_instanton_2022, heller_heavy-atom_2022, richardson_semiclassical_2015} 
The energy of an instanton trajectory can be obtained by taking the derivative of the action
\begin{equation}
    \frac{\partial S_n(x_\mathrm{i},x_\mathrm{f},\tau_n)}{\partial\tau_n}=E_n,
\end{equation}
where $\tau_0=\beta\hbar-\tau$ and $\tau_1=\tau$.
The stationary-action principle of instanton theory enforces energy conservation at the hopping point, such that $E_0=E_1\equiv E$.

The SCI rate in Eq.~\eqref{equ:instanton} can equivalently be written as\cite{richardson_semiclassical_2015,heller_instanton_2020} 
\begin{equation}
    k_\mathrm{SCI}Z_0=\sqrt{2\pi\hbar}\,\frac{\Delta^2}{\hbar^2}\sqrt{\frac{C_0C_1}{-\Sigma}}\,\mathrm{e}^{-S/\hbar}, \label{equ:instanton Sigma}
\end{equation}
\begin{equation}
    \Sigma=
    \begin{vmatrix}
        \frac{\partial^2 S}{\partial x'\partial x'} & \frac{\partial^2 S}{\partial x'\partial x''} & \frac{\partial^2 S}{\partial x'\partial\tau} \\
        \frac{\partial^2 S}{\partial x''\partial x'} & \frac{\partial^2 S}{\partial x''\partial x''} & \frac{\partial^2 S}{\partial x''\partial\tau} \\
        \frac{\partial^2 S}{\partial\tau\partial x'} & \frac{\partial^2 S}{\partial\tau\partial x''} & \frac{\partial^2 S}{\partial\tau^2}
    \end{vmatrix}, \label{equ:Sigma}
\end{equation}
where all square roots are combined into one and $\boldsymbol{\Sigma}$ has one negative eigenvalue.
Equation~\eqref{equ:instanton Sigma} will prove more useful for the analytic continuation to the non-convex regime, which we perform in the following section.
We thereby follow a similar approach to the analytic continuation of instanton theory to the inverted regime \cite{heller_instanton_2020}, where $\tau<0$. 
In that case, $\boldsymbol{C}_1$ is negative definite, $\boldsymbol{C}$ has $f$ negative eigenvalues, and $\boldsymbol{\Sigma}$ has $f+1$ negative eigenvalues (see the first two rows in Table \ref{tab:summary}). 
Despite these qualitative changes in several components of the prefactor, the total SCI rate expression [Eq.~\eqref{equ:instanton Sigma}] remains valid, which we will demonstrate to hold also in the non-convex regime. 

\red{For the practical application of instanton theory to multidimensional molecular systems, the ring-polymer framework is used, \cite{richardson_ring-polymer_2015, heller_instanton_2020, ansari_instanton_2022, fang_competing_2023, ansari_heavy-atom_2024, GRperspective}
where the instanton is discretized into $N$ replicas of the molecular geometry, or beads, along the path $\boldsymbol{x}\equiv\{x_1,\dots,x_N\}$, with $N_n$ equally spaced imaginary-time intervals $\epsilon_n\equiv\tau_n/N_n$ on each diabatic surface.
The action is then defined by 
\begin{align}
    \label{equ:RPaction}
    S_N(\boldsymbol{x},\tau)=& \sum_{i=1}^{N_0}\left(\frac{m\|x_{i}-x_{i-1}\|^2}{2\epsilon_0}+\epsilon_0\frac{V_0(x_{i-1})+V_0(x_{i})}{2}\right) \notag \\
    & +\sum_{i=N_0+1}^N\left(\frac{m\|x_{i}-x_{i-1}\|^2}{2\epsilon_1}+\epsilon_1\frac{V_1(x_{i-1})+V_1(x_{i})}{2}\right),
\end{align}
where cyclic boundary conditions $x_0\equiv x_N$ are imposed on the bead index $i$.
The instanton path can then be obtained by locating a saddle point of the ring-polymer action [Eq.~\eqref{equ:RPaction}] with respect to $\boldsymbol{x}$ and $\tau$, for which a variety of established algorithms are available.} \cite{richardson_ring-polymer_2015, heller_instanton_2020, ansari_instanton_2022, GRperspective}

\section{Instanton theory in the non-convex regime}

\subsection{Sign problem in the non-convex regime}

Previously, SCI has been applied to nonadiabatic processes where both diabatic PESs are convex near the crossing seam, as illustrated in Figure~\ref{fig:schematic}a. 
As explained in Section~\ref{sec:instanton}, for the convex case, $C_0$, $C_1$, $C$, and $-\frac{\mathrm{d}^2S}{\mathrm{d}\tau^2}$ should all be positive in the normal regime, while the signs of $C_1$ and $C$ should both be $\prn{-1}^f$ in the inverted regime.
However, if either of the two diabatic PESs exhibits a barrier near the crossing point, as illustrated in Fig.~\ref{fig:schematic}b, the surface becomes non-convex in that region, potentially altering the signs of the terms in the SCI prefactor.

In the following, we take $V_0$ to be non-convex and $V_1$ to be convex.  
The argument extends straightforwardly to cases where both surfaces are non-convex.
As a consequence, the diabatic Hessians along the instanton path on $V_0$ are no longer positive definite, in which case either the $\boldsymbol{C}$ or $\boldsymbol{C}_0$ matrix gains one extra negative eigenvalue, while $\frac{\mathrm{d}^2 S}{\mathrm{d}\tau^2}$ becomes positive (see Table~\ref{tab:summary} for a simple categorization and SI Section~S1 for a more in-depth discussion). 
We name this scenario the ``non-convex regime''.
We prove that a non-convex $V_0$ is the necessary condition for this regime to appear in one-dimensional systems and separable multidimensional systems (SI Section~S2), as well as in the classical limit for any system (Appendix~\ref{sec:classical}).
\blue{This suggests that similar behavior is expected in the general case.}

\renewcommand{\arraystretch}{1.2}
\begin{table}[!t]
    \centering
    \begin{tabular}{ccc|cc|c|c|c}
        \hline
        \multicolumn{3}{c|}{\multirow{2}{*}{Regime}} & \multirow{2}{*}{$\frac{C_0 C_1}{C}$} & \multirow{2}{*}{$\frac{\mathrm{d}^2 S}{\mathrm{d}\tau^2}$} & \multirow{2}{*}{$\tau$} & \multirow{2}{*}{\makecell{Curvature of $V_n$\\along instantons}} & \multirow{2}{*}{Schematic} \\ 
         &  &  &  &  &  & \\
        \hline
        \multicolumn{2}{c|}{\multirow{4}{*}{Convex}} & \multirow{2}{*}{Normal} & \multirow{2}{*}{$+$} & \multirow{2}{*}{$-$} & \multirow{2}{*}{$+$} & \multirow{4}{*}{Dominantly positive} & \multirow{2}{*}{\includegraphics[height=0.42in]{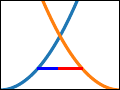}} \\
        \multicolumn{2}{c|}{} &  &  &  &  &  & \\
        \cline{3-3}
        \multicolumn{2}{c|}{} & \multirow{2}{*}{Inverted} & \multirow{2}{*}{$+$} & \multirow{2}{*}{$-$} & \multirow{2}{*}{$-$} &  & \multirow{2}{*}{\includegraphics[height=0.42in]{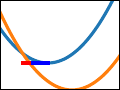}} \\
        \multicolumn{2}{c|}{} &  &  &  &  &  & \\
        \cline{3-3}
        \hline
        \multicolumn{2}{c|}{\multirow{4}{*}{\makecell{Non-\\convex}}} & \multirow{2}{*}{Normal} & \multirow{2}{*}{$-$} & \multirow{2}{*}{$+$} & \multirow{2}{*}{$+$} & \multirow{4}{*}{\makecell{Balanced positive/negative \\ or dominantly negative}} & \multirow{2}{*}{\includegraphics[height=0.42in]{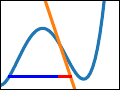} \includegraphics[height=0.42in]{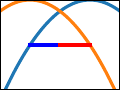}} \\
        \multicolumn{2}{c|}{} &  &  &  &  &  & \\
        \cline{3-3}
        \multicolumn{2}{c|}{} & \multirow{2}{*}{Inverted} & \multirow{2}{*}{$-$} & \multirow{2}{*}{$+$} & \multirow{2}{*}{$-$} &  & \multirow{2}{*}{\includegraphics[height=0.42in]{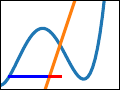} \includegraphics[height=0.42in]{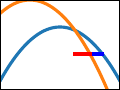}} \\
        \multicolumn{2}{c|}{} &  &  &  &  &  & \\
        \hline
    \end{tabular}
    \caption{Summary of instantons in the convex and non-convex regimes.}
    \label{tab:summary}
\end{table}

\begin{figure}[t]
    \centering
    \subfigure{\includegraphics[width=3.18in]{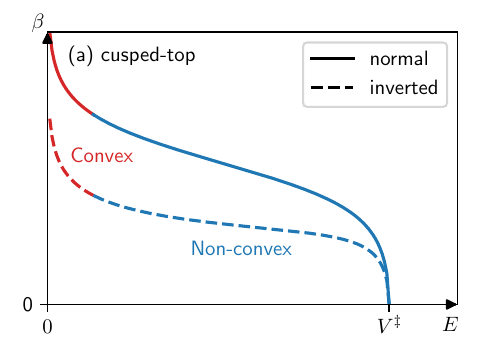}}
    \subfigure{\includegraphics[width=3.27in]{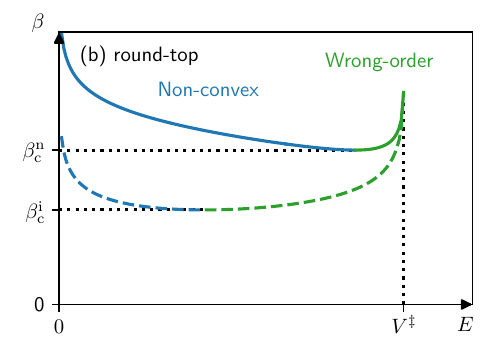}}
    \caption{Illustration of inverse temperature $\beta$ (imaginary time) of (a) cusped-top instantons (upper path in Figure~\ref{fig:schematic}b) and (b) round-top instantons (lower path in Figure~\ref{fig:schematic}b) as a function of tunneling energy $E$. 
    The solid line is for the normal regime and the dashed line is for the inverted regime. 
    Both regimes have the same crossover point. 
    The crossing point energy $V^\ddagger$ and the crossover inverse temperature $\beta_\mathrm{c}$ of round-top instantons for each case are marked on the axes. 
    $\beta_\mathrm{c}$ corresponds to the highest temperature where a round-top instanton can exist, and the superscripts n and i stands for the normal and inverted regimes respectively. 
    The wrong-order instantons have an overall minus sign in their prefactors, leading to unphysical imaginary rates.}
    \label{fig:beta-E}
\end{figure}

Instantons in the non-convex regime exhibit significant diversity, making it important to establish a clear categorization.
As in convex systems, reactions in non-convex systems can also be classified into Marcus normal and inverted regimes, depending on whether the reactant and product lie on opposite sides or the same side of the crossing seam.
Similarly, the distinction between normal-regime and inverted-regime instantons  based on whether $\tau>0$ or $\tau<0$, \cite{richardson_semiclassical_2015, heller_instanton_2020} remains valid regardless of whether the system is in the convex or non-convex regime.

A straightforward and practical categorization of non-convex instantons is based on the shape of the potential-energy curves along the instanton paths and the nature of their tunneling mechanism, which divides them into ``round-top'' and ``cusped-top'' types.
Specifically, the round-top instantons simultaneously traverse a potential energy barrier and switch electronic state, representing the concerted mechanism (lower nonadiabatic instanton in Figure~\ref{fig:schematic}b).
The cusped-top instantons switch electronic state near a potential energy barrier, but do not traverse it.  
They can represent one step of the sequential mechanism (upper nonadiabatic instanton in Figure~\ref{fig:schematic}b). 
The cusped-top non-convex instantons share many similarities with convex-regime instantons, which are also cusped-topped. \cite{GRperspective}
For example, as the temperature increases, it continuously collapses towards the MECP (Figure~\ref{fig:beta-E}a).
In fact, as the temperature decreases, a non-convex instanton may even transition to a convex instanton (Figure~\ref{fig:beta-E}a), for the reason that the majority of the instanton imaginary time is spent near the minimum rather than the barrier.

Round-top instantons, however, exhibit distinct behavior.
A key feature of them is that they only exist below a certain temperature (we refer to it as the crossover temperature $T_\text{c}$), which is a property normally only found for BO instantons. 
Note that it is difficult to predict $T_\text{c}$ for the round-top instanton, since unlike BO instantons, it is not simply determined by the imaginary frequency at the barrier top.
Nevertheless, we find that it is typically higher than the $T_\text{c}$ of a BO instanton on $V_0$.
This is because the barrier formed by crossing PESs is narrower than the single-surface barrier (Figure~\ref{fig:schematic}b), making it easier for an instanton to exist.
The unusual appearance of $T_\text{c}$ in nonadiabatic instantons arises from the concerted instanton traversing a rounded barrier, similar to a Born--Oppenheimer (BO) instanton.
In contrast, convex instantons exist at any temperature. 

Another key feature is that as the energy $E$ of a round-top instanton increases, it undergoes a transition to a `wrong-order' instanton (Figure~\ref{fig:beta-E}b).
The wrong-order instanton has an extra negative eigenvalue in the prefactors, resulting in an imaginary instanton partition function.
We note that identical behavior has been previously discovered in BO instantons on a broad-top barrier \cite{fang_simultaneous_2017}.
The wrong-order instanton cannot be directly used to compute canonical rates, as Eq.~\eqref{equ:instanton Sigma} will give an unphysical imaginary rate constant.

\subsection{Analytic continuation}

The sign changes observed in various terms of the SCI prefactor in the non-convex regime may appear problematic.
However, Eq.~\eqref{equ:instanton Sigma} remains well-defined even if we relax the sign constraints on the individual prefactor terms imposed by the derivation, as can be shown through the following analytic continuation.

\begin{figure}[!b]
    \centering
    \includegraphics[width=\linewidth]{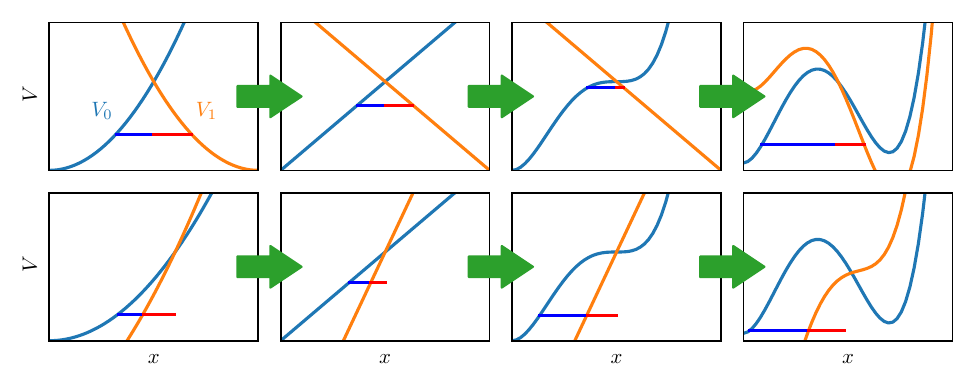}
    \caption{Illustration of an analytic continuation scheme from a convex regime instanton to a non-convex regime instanton. 
    The tunneling paths on either state (in blue and red) form the instantons.
    The top row is for the normal regime, and the bottom row is for the inverted regime.
    From left to right, we deform the potentials from a 1D convex model to a linear model, and then introduce negative curvature in $V_0$ and $V_1$.}
    \label{fig:scheme}
\end{figure}

We propose an analytic continuation scheme via the deformation of the potential to cover all types of non-convex-regime instantons.
It is well established that instanton theory performs very well in both the normal and inverted regimes of convex systems, such as those illustrated in the first column of Figure \ref{fig:scheme})\cite{heller_instanton_2020}. 
Then for any given non-convex instanton at a near-barrier crossing (e.g., the final column in Figure \ref{fig:scheme}), 
one can construct a continuous deformation of the PESs such that the instanton smoothly evolves from the convex regime to the target non-convex configuration.

\blue{Since the instanton rate is known to be asympotically correct in the convex regime, it must remain correct as we continuously deform the parameters of the system as long as the expression remains analytic, i.e., does not encounter any singularies.}
A potentially problematic case arises in systems with linear potentials, where $C\to0$ and $\frac{\mathrm{d}^2 S}{\mathrm{d}\tau^2}\to\infty$ \cite{richardson_semiclassical_2015}. 
However, as shown in Ref. \citenum{richardson_semiclassical_2015}, the instanton rate remains well-defined and, in fact, exactly reproduces FGR for linear potentials. 
In this case, the singularity in $\frac{\mathrm{d}^2 S}{\mathrm{d}\tau^2}$ cancels with the vanishing $C$, yielding a finite result. 
Equivalently, one finds that $\Sigma$ remains well-defined. 
In this way, we can justify the analytical continuation to the regime where $C$ and $\frac{\mathrm{d}^2 S}{\mathrm{d}\tau^2}$ swap their signs, which is the most common non-convex case.
Further proofs for the case regarding conjugate points can be found in SI Section~S3.

Finally, we present additional evidence supporting the validity of the analytically continued instanton rate expression in the non-convex regime.
Firstly, in the classical limit, the instanton rate correctly reduces to the NA-TST rate in the non-convex regime (proof given in Appendix \ref{sec:classical}).
Secondly, we provide a direct first-principles derivation of instanton theory in the Hamilton--Jacobi formalism, \cite{richardson_semiclassical_2015} based on Green's functions instead of propagators for the non-convex normal regime.
In particular, we prove in SI Section~S4 that Green's functions remain well behaved even in situations where propagators become problematic, and for all systems tested thus far, the resulting prefactors are consistently positive, thereby justifying the steepest-descent integration.
Further, we use the Green's function formulation to show that well-behaved semiclassical propagators for the individual paths can still be derived when conjugate points are encountered. 
Although this derivation does not extend naturally to the inverted regime, where Green's functions become highly oscillatory and lack a well-defined peak \cite{heller_instanton_2020}, 
it provides further independent support for our analytic continuation approach.

\section{Numerical results}
\subsection{Benchmarking non-convex instanton theory in model systems}
\label{subsec:benchmark}

\begin{figure}[!t]
    \centering
    \includegraphics[width=\linewidth]{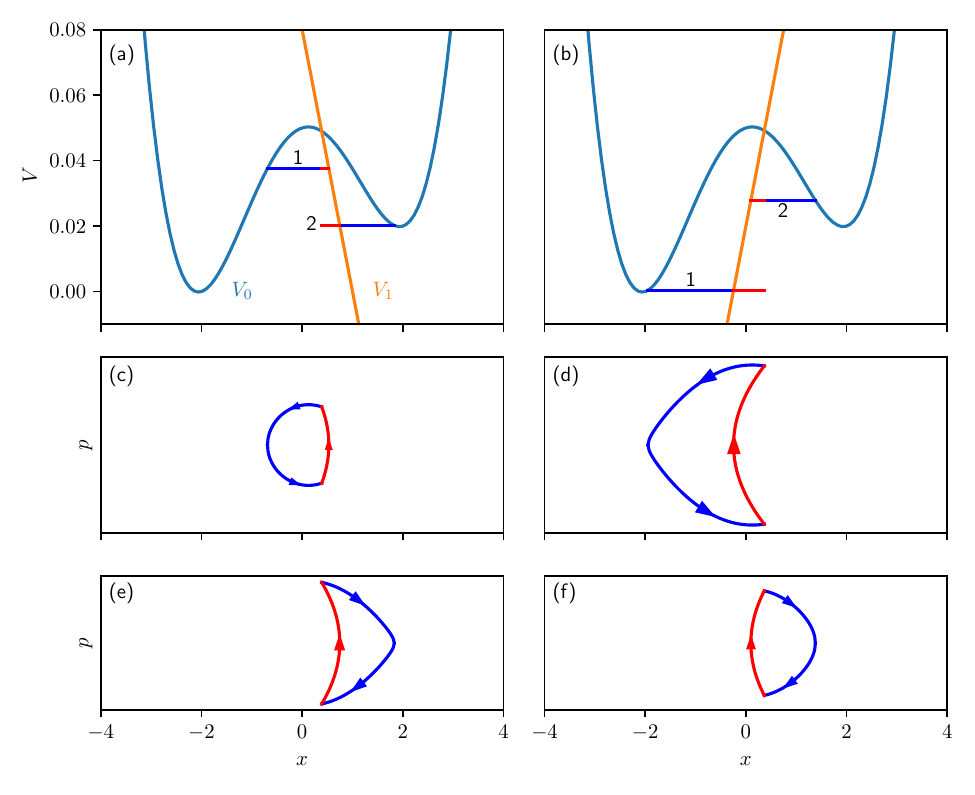}
    \caption{
    Diabatic potentials defined in Eq.~\eqref{equ:modelpots}, where the instanton describing concerted tunneling is in the normal regime for model A (a) and in the inverted regime for model B (b). 
    All instantons displayed are optimized at $\beta=1000$. 
    Panels c-f show the instantons in phase space.
    (c) Instanton 1 in Panel a (non-convex). (d) Instanton 1 in Panel b (non-convex). (e) Instanton 2 in Panel a (convex). (f) Instanton 2 in Panel b (non-convex).
    Note that there is no fundamental difference in the shapes of the convex and non-convex instantons in phase space. 
    }
    \label{fig:models}
\end{figure}

In this section, we benchmark our extended instanton theory using a one-dimensional model system composed of an asymmetric double well intersected by a linear potential, defined by
\begin{subequations}
\label{equ:modelpots}
\begin{align}
    V_0 &=\lambda\left(\frac{x}{\xi}+1\right)^2\left(\frac{x}{\xi}-1\right)^2+\kappa_0 x+\varepsilon_0,\\
    V_1 &=\kappa_1x+\varepsilon_1,
\end{align}
\end{subequations}
with $m=1836$, $\lambda=0.04$, $\xi=2$, $\kappa_0=0.005$, $\varepsilon_0=0.01$ and $\Delta=0.0004$ given in atomic units. 
We test two parameter sets for $\kappa_1$ and $\varepsilon_1$  to cover both the non-convex normal and inverted regime, as shown in Figure \ref{fig:models}. 
We set $\kappa_1=-0.08$, $\varepsilon_1=0.08$ for the normal-regime model A, and $\kappa_1=0.08$, $\varepsilon_1=0.02$ for the inverted-regime model B. 
Figure \ref{fig:models} illustrates the model potentials along with two optimized nonadiabatic instantons at $\beta=1000$ for each model, where the \hl{round-top} instanton is labeled as 1 and the \hl{cusped-top} instanton is labeled as 2. 

\begin{figure}[!t]
    \centering
    \subfigure{\includegraphics[width=3.48in]{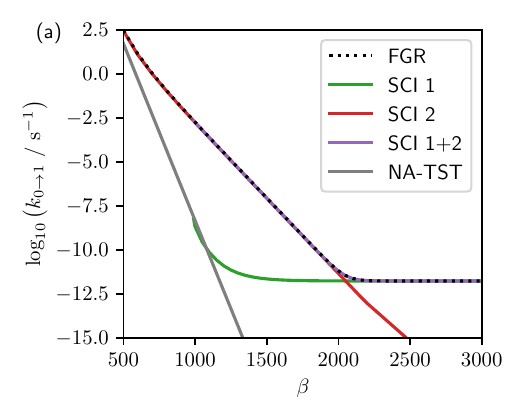}}
    \subfigure{\includegraphics[width=2.97in]{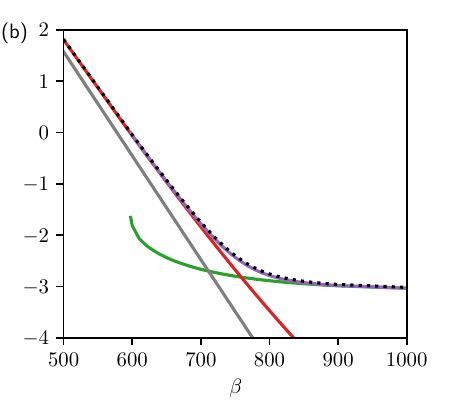}}
    \caption{Rate constants for the reaction from state 0 to state 1 $k_{0\to1}$ as a function of inverse temperature $\beta$ for model A (a) and model B (b). 
    \hl{SCI 1 and 2 represent the rates given by instantons 1 and 2 in Figure}~\ref{fig:models}.
    \hl{SCI 1+2 stands for the sum of the two rates.}
    For FGR, the exact reactant partition function is used, while the remaining methods employ the harmonic approximation to $Z_0$ evaluated at the minimum of the left well.} 
    \label{fig:rates}
\end{figure}
In Figure \ref{fig:rates}, we present the rate constants for the transitions from state 0 to state 1 in both models, computed with SCI, NA-TST and, for comparison, FGR. 
In the FGR calculations, we use the exact reactant partition function $Z_0$, obtained by summing the Boltzmann factors associated with the vibrational eigenstates of the reactant potential.
Instead, the SCI and NA-TST calculations evaluate $Z_0$ based on a harmonic approximation around the minimum of the left well, even for instanton 2, which describes transitions starting from the right well (see Figure~\ref{fig:models}).
The reason is that this yields a rate $k_2$ equivalent to the pre-equilibrium approximation, which assumes rapid equilibration between the left and right wells before the electronic state transition, and makes it directly comparable to $k_1$ from instanton 1 and the FGR rate.
Figure \ref{fig:rates} highlights the excellent agreement between the FGR rate and the SCI rate, with the latter obtained as the sum of contributions from instantons 1 and 2. 
The maximum relative error is only $4\%$ for the first model and $7\%$ for the second. 

As expected, the \hl{cusped-top} instanton dominates at high temperatures, where it approaches the NA-TST rate, whereas instanton 1 ceases to exist above its crossover temperature. 
At low temperatures, however, this situation changes dramatically, with the \hl{round-top} instanton 1 becoming dominant. 
This is because at low temperature, only the lowest energy levels in the left well are significantly populated. 
These asymptotic trends confirm that instanton 1 is a purely quantum-mechanical deep-tunneling path, while instanton 2 transitions smoothly from deep to shallow tunneling, and ultimately to the classical limit.

\subsection{Nonadiabatic tunneling mechanisms in the non-convex regime}
\label{subsec:mechanisms}

In practice, we are often interested in the overall rate, %
\blue{starting from the left reactant well of Fig.~\ref{fig:models} and ending on the product state.\footnote{Alternatively, the nonadiabatic transition can preeced the adiabatic step, which would be appropriate for the system depicted in Fig.~\ref{fig:schematic}(b).}}
There are two available mechanisms for such a process, with one being sequential and the other being concerted.
The sequential mechanism, where the system first crosses the barrier on $V_0$ and subsequently undergoes a nonadiabatic transition to $V_1$ via instanton 2 in Figure \ref{fig:models}, was previously thought to be the only pathway for the overall reaction. \cite{harvey_spin-forbidden_2014}
In the high-temperature limit, the tunneling rate associated with this mechanism reduces to its classical counterpart. 
In contrast, the concerted mechanism, in which both processes occur concomitantly, represents a deep-tunneling pathway with no classical analogue.
In the following we demonstrate that this purely quantum-mechanical pathway can in fact dominate over the sequential mechanism at low temperatures. 

The evaluation of the overall reaction rate for the sequential mechanism involves both the adiabatic rate on the reactant state and the nonadiabatic rate for hopping from the right reactant well to the product state.
The rate constant for the adiabatic step, traversing the diabatic barrier without a change in electronic state, denoted as step 3, can be accurately described by Born--Oppenheimer instanton theory \cite{richardson_ring-polymer_2018}.
The concerted rate $k_\mathrm{con}$ is just $k_1$, while the sequential rate is a combination of steps 2 and 3, given by
\begin{equation}
   k_\mathrm{seq}=\frac{k_2}{2 k_{-3}}\left(k_3-k_{-3}-k_2+\sqrt{\left(k_3-k_{-3}-k_2\right)^2+4 k_3 k_{-3}}\right), \label{equ:sequential}
\end{equation}
where $k_{-3}$ is the rate constant of the reverse reaction of step 3, and, in contrast to the previous subsection, $k_2$ is calculated with $Z_0$ evaluated in the right well. 
Equation \eqref{equ:sequential} is derived by analytically solving the rate equations under an infinite-sink approximation on the product side to obtain the time-dependent rate $k\left(t\right)$, and then taking the limit as $t\to+\infty$ (see Appendix~\ref{sec:overall rate} for the derivation). 
If $k_2\gg \{k_3,k_{-3}\}$, the overall rate $k_\mathrm{seq}$ reduces to the rate-determining step $k_3$. 
Conversely, if $k_2\ll \{k_3,k_{-3}\}$, the rate-determining step becomes $k_2$, and $k_\mathrm{seq}$ reduces to the pre-equilibrium approximation $k_2\frac{k_3}{k_{-3}}$.

\begin{figure}[!b]
    \centering
    \subfigure{\includegraphics[width=3.37in]{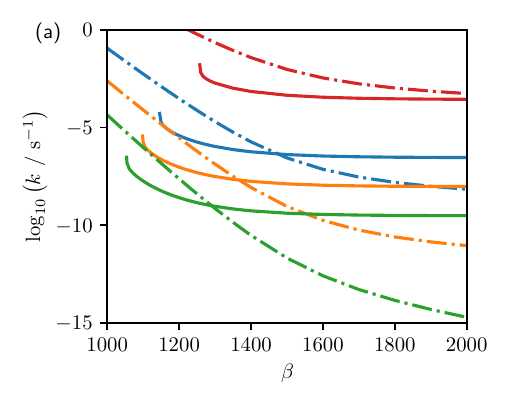}}
    \subfigure{\includegraphics[width=3.08in]{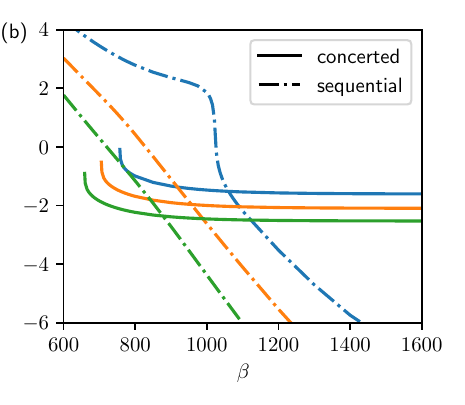}}
    \caption{Overall rate constants $k$ for the sequential and concerted mechanisms in model A (a) and model B (b) as functions of inverse temperature $\beta$.
    Each color corresponds to a different value of $\kappa_0$: red for $\kappa_0=-0.006$, blue for $\kappa_0=-0.002$, orange for $\kappa_0=0$ and green for $\kappa_0=0.002$. Solid and dashed-dotted lines of the same color represent the concerted and sequential contributions, respectively.
    }
    \label{fig:mechanisms}
\end{figure}

We explore the competition between sequential and concerted mechanisms by studying models A and B from the previous subsection (see Figure~\ref{fig:models}) across a range of different parameters. 
The relative importance of the two mechanisms depends on several factors, including the potential-energy difference between the reactant wells, the diabatic coupling, the distance between the TS and the MECP, the length of the reaction path, and the barrier height. 
Here, we focus on the effect of the energy difference between the reactant wells, which is controlled by $\kappa_0$ (Figure \ref{fig:mechanisms}), while keeping all other parameters fixed as defined in Section~\ref{subsec:benchmark}.
Since the potentials used here qualitatively mimic the typical reaction scenario at near-barrier crossings, we expect the qualitative trends to carry over to more complex molecular systems.
In fact, the theoretical framework in this paper has recently been used to explore the concerted tunneling pathway in carbenes, where the potential energy profiles are rather similar to the model used here. \cite{carbenes}

In Figure~\ref{fig:mechanisms}, we consider exothermic ($\kappa_0<0$), symmetric ($\kappa_0=0$) and endothermic ($\kappa_0>0$) adiabatic reaction steps. 
As expected from the results in Section~\ref{subsec:benchmark}, the sequential mechanism dominates at high temperatures, while the concerted mechanism governs the low-temperature regime. 
It can be seen that the temperature at which the two rates coincide increases with growing endothermicity of the adiabatic reaction step, i.e., the concerted mechanism dominates already at higher temperatures.
For $\kappa_0=-0.006$, the crossing point is at a far lower temperature and even beyond the shown temperature range. 
This can be explained by the fact that a more negative $\kappa_0$ leads to not only a more exothermic reaction but also a narrower barrier, which causes $k_\mathrm{seq}$ to increase. 

Note that \blue{there is no requirement for a crossover to occur at all, as} the sequential mechanism can be faster than the concerted one at all temperatures. %
As $\kappa_0$ is varied from negative to positive in model A, $k_\mathrm{seq}$ decreases much more rapidly at low temperatures.
\blue{The reason is that, for $\kappa_0>0$, the instanton energy is bounded from below by the minimum of the right well, and thus the reaction always requires an initial thermal activation, leading to an exponential suppression of the rate constant.
In contrast, for $\kappa_0\le0$, the instanton energy approaches the minimum of the left well, corresponding to a tunneling process from the bottom of the well, which yields a finite rate constant in the low-temperature limit.}

Model B exhibits similar trends for the competition between the two mechanisms. 
However, while in model A, $k_\mathrm{seq}\approx k_3$, model B requires the usage of Eq.~\eqref{equ:sequential} because the rate constants entering the formula for the sequential rate are now comparable in magnitude.
As the temperature decreases, the dominant contribution to $k_\mathrm{seq}$ shifts from $k_2\frac{k_3}{k_{-3}}$ to $k_3$, reflecting a change in the rate-determining step from $k_2$ to $k_3$.
This transition is especially pronounced for $\kappa_0=-0.002$ (dashed-dotted blue curve in Figure~\ref{fig:mechanisms}b), where a sharp drop in the rate occurs around $\beta=1000$.
The origin of this behavior is that $k_3\gg k_{-3}$ for this parameter set. 
In that regime, when $k_2$ becomes comparable to $k_3$, the term $k_2\frac{k_3}{k_{-3}}$ grows much larger than $k_3$, resulting in a sudden decrease in $k_\mathrm{seq}$ as it transitions from being governed by $k_2\frac{k_3}{k_{-3}}$ to being limited by $k_3$ alone. This illustrates the rich interplay between different reaction steps that can emerge from varying a single parameter, and underscores the importance of capturing competing nuclear quantum effects in non-convex molecular systems in their full complexity, as is made possible by our extension of instanton theory to this scenario. 

We have also investigated intersecting double-well models similar to the one shown in Figure~\ref{fig:schematic}b. As demonstrated in SI Section~S5, these models exhibit no qualitative differences from our models A and B, which is why we have focused our analysis on those two representative cases.

\subsection{Isomerization of HON}

\begin{figure}[!b]
    \centering
    \subfigure{\includegraphics[width=3.2in]{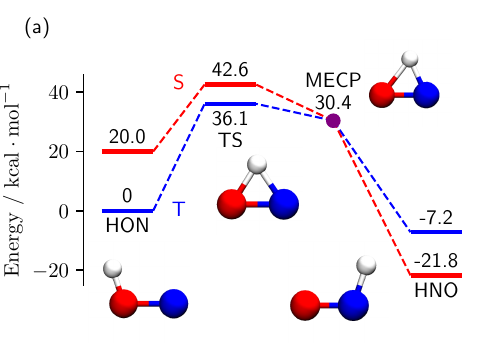}}
    \subfigure{\includegraphics[width=3.2in]{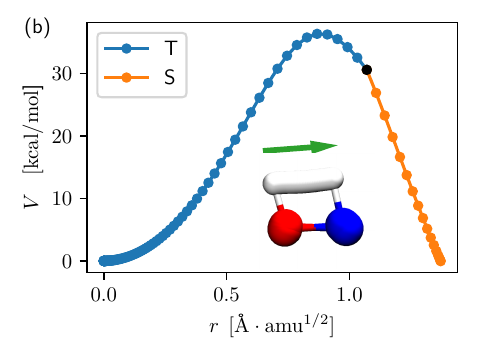}}
    \caption{\hl{(a) The energy diagram and (b) the concerted instanton path at 125 K from triplet HON to singlet HNO at the level of UDSD-PBEP86/def2-TZVPP.}}
    \label{fig:HON energy}
\end{figure}

\hl{In this section, we apply instanton theory to the isomerization of HON to HNO.
This isomerization can take place in solid argon at 10 K after irradiation,} \cite{maier_isonitroso_1999} \hl{while its spontaneous reaction has not been reported.
This system is probably the simplest real molecular system of near-barrier spin-crossover, and we thus use this system to demonstrate the competition between the two mechanisms.

The \textit{ab initio} calculations are computed on-the-fly by double-hybrid density functional theory (DFT), using the UDSD-PBEP86 functional}\cite{kozuch_dsd-pbep86_2011}\hl{ with the def2-TZVPP basis set}\cite{weigend_balanced_2005, weigend_accurate_2006}\hl{ as implemented in Gaussian 16.}\cite{frisch_gaussian_2016}
\hl{To obtain the spin-orbit coupling (SOC) $\Delta$, we employ the complete active space self-consistent field method (CASSCF) at the level of CASSCF(12,12)/cc-pVTZ using ORCA 6.1.} \cite{RN178, RN204, RN243, RN269}

\begin{figure}[!b]
    \centering
    \includegraphics[width=0.5\linewidth]{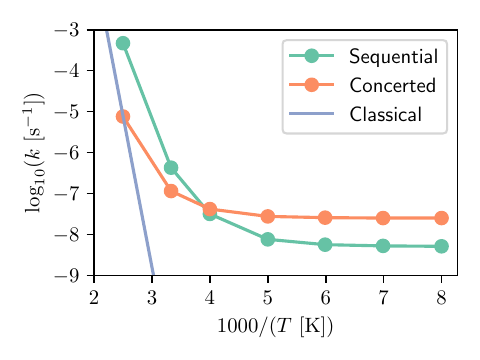}
    \caption{\hl{Overall rate constants $k$ for the sequential and concerted mechanisms from triplet HON to singlet HNO as functions of temperature. The classical counterpart of the sequential mechanism is also plotted for comparison.}}
    \label{fig:HON rates}
\end{figure}

\hl{As shown in Figure}~\ref{fig:HON energy}\hl{a, the isomerization reaction of a triplet HON should first overcome a diabatic barrier of 36.1 kcal/mol on the triplet state, and then cross from the triplet to singlet state to obtain a singlet HNO, which is quantitatively consistent with previously reported calculations.} \cite{guadagnini_global_1995}
\hl{The SOC evaluated at the MECP is 81.6 cm$^{-1}$. 
Two competing tunneling mechanisms involved in this system are similar to that of model A used in Sections.}~\ref{subsec:benchmark}\hl{ and }\ref{subsec:mechanisms}
\hl{Since the energy of singlet HON is significantly higher triplet HON, we do not consider reactions starting from the singlet state.

The aforementioned sequential and concerted tunneling mechanisms can be found in using ring-polymer instanton optimization, with the latter indicated by the instanton path shown in Figure}~\ref{fig:HON energy}\hl{b.
Note that $k_\mathrm{seq}$ approximately equals to the BO instanton rate of the first step of the sequential mechanism, since the second step is markedly faster. 
The competition between the two mechanisms is revealed in Figure~}\ref{fig:HON rates}\hl{.
Consistent with the results in Section~}\ref{subsec:mechanisms}\hl{, concerted tunneling dominates over the sequential process at temperatures lower than around 250 K, while the sequential pathway becomes the major mechanism at higher temperatures. 
The classical rate on the other hand is always lower than the sequential rate at below 400 K, suggesting that tunneling plays a key role in this reaction even at such high temperatures.
The slow rate of $2.52\times10^{-8}~\mathrm{s}^{-1}$ at low temperature is also consistent with the experimental fact that the spontaneous isomerization was not observed on the experimental timescale.} \cite{maier_isonitroso_1999}

\section{Conclusions}

We have developed an extended semiclassical instanton theory capable of describing nonadiabatic tunneling in systems where an electronic-state crossing occurs near a diabatic barrier. 
Unlike previous instanton formulations that assumed mostly convex diabatic potentials, the present approach generalizes SCI to this non-convex regime, where tunneling paths may traverse large regions of negative curvature.
The fluctuation factors of the non-convex instantons exhibit different signs than their convex counterparts due to the occurrence of conjugate points along the paths, which appear to invalidate the theory. 
We resolve this issue by analytically continuing the instanton rate expression to the non-convex regime.
Benchmark comparisons to other methods for model systems demonstrate that the non-convex instanton theory yields excellent agreement with fully quantum-mechanical FGR rates across a wide range of parameters, including in the deep-tunneling regime, and reduces to NA-TST in the classical limit. 

It is particularly interesting that the non-convex instanton theory enables the study of the competition between a sequential reaction mechanism, and a deep-tunneling concerted mechanism without classical analogue. 
Previously, only the sequential mechanism involving an intermediate adiabatic step followed by a nonadiabatic transition has been considered. 
While this reaction path prevails at high temperatures, we show that the concerted mechanism can dominate at low temperatures. 
The interplay between the two mechanisms strongly depends on a multitude of factors, such as the potential energy asymmetry, diabatic coupling strength, and barrier shape. 
A rich dynamical behavior is thus observed, emphasizing the need for a unified framework, as provided by our extended instanton theory, that captures the full complexity of nonadiabatic quantum tunneling in molecular systems.

In addition, our extended instanton theory provides a computationally efficient and quantitatively accurate tool for studying competing sequential and concerted nonadiabatic tunneling in molecular systems with near-barrier crossings. 
\hl{We apply this theory to the isomerization of HON, revealing the competitive tunneling consistent with those observed in model studies.}
This was also recently demonstrated in \Refx{carbenes}, where one of the instantons discussed here was used to explain the temperature-dependent selectivity observed in reactions of triplet carbenes.

\begin{acknowledgement}

Z.Y. and W.F. are supported by the National Natural Science Foundation of China under Grant No. 22321003.
E.R.H is grateful for financial support from the Swiss National Science Foundation through Grant 214242.
J.O.R. acknowledges financial support from the Swiss National Science Foundation through Project 207772.

\end{acknowledgement}

\begin{suppinfo}

Supporting Information: Additional discussion on the categorization of non-convex instantons; Detailed derivations for necessary conditions for the non-convex regime and singularity cancellation; Supporting numerical calculations for impact of model parameters on rate constant behavior.

\end{suppinfo}

\begin{appendices}

\section{Classical limit}
\label{sec:classical}
\setcounter{equation}{0}
\renewcommand{\theequation}{\thesection\arabic{equation}}

As shown in SI Section~S2, an extra minus may arise in $C_n$ due to the existence of a conjugate point.
However, in the classical limit, $C_n$ will never encounter this problem, since the conjugate point can only exist when a trajectory is long enough.\cite{landau_mechanics_2005}
However, we will show that $\boldsymbol{C}$ may have negative eigenvalues in the classical limit of the non-convex regime. 

The classical short-time limit of the action takes the form of \cite{richardson_semiclassical_2015}
\begin{equation}
    S_n=\frac{m\left\|x_-\right\|^2}{2z_n}+V_n\left(x_+\right)z_n, \label{equ:short-time action}
\end{equation}
where $x_+=\frac{1}{2}\left(x'+x''\right),x_-=x'-x''$, and $\boldsymbol{C}$ is block-diagonal with entries
\begin{equation}
    \frac{\partial^2S}{\partial x_+\partial x_+}=\beta\hbar\widetilde{\boldsymbol{H}}, \label{equ:C+ cl}
\end{equation}
\begin{equation}
    \frac{\partial^2S}{\partial x_-\partial x_-}=\frac{m\beta\hbar}{\tau_0\tau_1}\boldsymbol{I}_f, \label{equ:C- cl}
\end{equation}
where $\widetilde{\boldsymbol{H}}\equiv\frac{1}{\beta\hbar}\left(\tau_0\boldsymbol{H}_0+\tau_1\boldsymbol{H}_1\right)$, and $\boldsymbol{I}_f$ is an $f\times f$ identity matrix. 
\blue{Neither of these expressions is necessarily positive definite.}
Similar to the 1D system, the sign of Eq.~\eqref{equ:C- cl} indicates whether it is in the normal or inverted regime, %
and imaginary frequencies can appear in $\boldsymbol{H}_0$ and $\boldsymbol{H}_1$, meaning that they also appear in $\widetilde{\boldsymbol{H}}$, and hence in $\frac{\partial^2S}{\partial x_+\partial x_+}$.

However, it can be proven that instanton theory does not break down in the classical limit even if the positive-definiteness of $\boldsymbol{C}$ is not satisfied.
Inserting Eq.~\eqref{equ:short-time action} into Eq.~\eqref{equ:semiclassical rate}, we obtain
\begin{equation}
    k_\mathrm{cl}Z_0=\frac{1}{\hbar^2\left(2\pi\hbar\right)^f}\iiint\left|\Delta\right|^2\left(\frac{m}{\sqrt{z_0z_1}}\right)^{f}\mathrm{e}^{-\frac{m\beta}{2z_0z_1}\left\|x_-\right\|^2-\frac{1}{\hbar}\left[V_0\left(x_+\right)z_0+V_1\left(x_+\right)z_1\right]}\,\mathrm{d}x_-\mathrm{d}x_+\mathrm{d}t, \label{equ:cl rate (start)}
\end{equation}
and $x_-$ can be easily integrated out analytically to give 
\begin{equation}
    k_\mathrm{cl}Z_0=\frac{1}{\hbar^2}\left(\frac{m}{2\pi\beta\hbar^2}\right)^{f/2}\iint\left|\Delta\right|^2\mathrm{e}^{-\frac{1}{\hbar}\left[V_0\left(x_+\right)z_0+V_1\left(x_+\right)z_1\right]}\,\mathrm{d}x_+\mathrm{d}t. \label{equ:cl rate (step 1)}
\end{equation}
Next, we would integrate Eq.~\eqref{equ:cl rate (step 1)} over $x_+$ first and then $t$ if we follow the previous derivation of instanton theory.
However, the potential non-positive-definiteness of Eq.~\eqref{equ:C+ cl} may prevent us from doing so. 
Nevertheless, we can alternatively integrate Eq.~\eqref{equ:cl rate (step 1)} over $t$ first by Fourier transform to give
\begin{equation}
    k_\mathrm{cl}Z_0=\frac{2\pi}{\hbar}\left(\frac{m}{2\pi\beta\hbar^2}\right)^{f/2}\int\left|\Delta\right|^2\mathrm{e}^{-\beta\widetilde{V}\left(x_+\right)}\delta\left[V_0\left(x_+\right)-V_1\left(x_+\right)\right]\,\mathrm{d}x_+, \label{equ:cl rate (step 2)}
\end{equation}
where $\widetilde{V}\equiv\frac{1}{\beta\hbar}\left(\tau_0V_0+\tau_1V_1\right)$. 
After steepest-descent approximation applied to Eq.~\eqref{equ:cl rate (step 2)}, the commonly used classical golden-rule rate is reproduced as
\begin{equation}
    k_\mathrm{cl}Z_0=\sqrt{\frac{2\pi m}{\beta\hbar^2}}\frac{\Delta^2}{\hbar\left|\boldsymbol{g}_0^\ddagger-\boldsymbol{g}_1^\ddagger\right|}Z^\ddagger\,\mathrm{e}^{-\beta V^\ddagger},
\end{equation}
\begin{equation}
    Z^\ddagger=\prod_{i=1}^{f-1}\frac{1}{\beta\hbar\widetilde{\omega}_i},
\end{equation}
where $\boldsymbol{g}_n^\ddagger$ are gradients, $\widetilde{\omega}_i$ are vibrational frequencies obtained by diagonalizing $\widetilde{\boldsymbol{H}}$ (after projecting out the reaction coordinate defined by the direction of the gradient), and all quantities are evaluated at the MECP.

We find from the derivation above that starting from Eq.~\eqref{equ:cl rate (start)}, the only approximation we make is the steepest-descent approximation, requiring all $\widetilde{\omega}_i$ (other than the frequency in the direction of the reaction coordinate) to be real. 
This condition can always be satisfied since it is also the necessary condition for a point to be MECP, and the imaginary frequencies (if any), typically associated with the reaction coordinate will always be projected out. 
Therefore, instanton theory reduces to the correct classical rate, regardless of the definiteness of $\widetilde{\boldsymbol{H}}$.

\section{Derivation of the overall reaction rate}
\label{sec:overall rate}
\setcounter{equation}{0}

\begin{figure}[t]
    \centering
    \begin{tikzpicture}[node distance = 2 cm, thick, main/.style = {draw, circle}] 
        \node[main] (1) {R};
        \node[main] (2) [above right of=1] {I};
        \node[main] (3) [below right of=2] {P};
        \draw[->] ([xshift=-0.5 mm, yshift=0.5 mm]1.north east) -- ([xshift=-0.5 mm, yshift=0.5 mm]2.south west) node[midway, above, sloped]{$k_3$}; 
        \draw[<-] ([xshift=0.5 mm, yshift=-0.5 mm]1.north east) -- ([xshift=0.5 mm, yshift=-0.5 mm]2.south west) node[midway, below, sloped]{$k_{-3}$};
        \draw[->] (1) -- (3) node[midway, below]{$k_1$};
        \draw[->] (2) -- (3) node[midway, above, sloped]{$k_2$};
    \end{tikzpicture}
    \caption{A schematic reaction network illustrating the systems in Figure \ref{fig:models}. There are two pathways leading from the reactant (R) to the product (P), with one path being direct and the other through an intermediate species (I).}
    \label{fig:graph}
\end{figure}
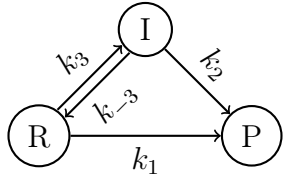

Consider a multi-step reaction network system as illustrated in Figure \ref{fig:graph}. The reaction kinetics can be described using the law of mass action as
\begin{align}
    \deriv{1}{\conc{R}}{t}&=-\prn{k_1+k_3}\conc{R}+k_{-3}\conc{I}, \label{equ:R} \\
    \deriv{1}{\conc{I}}{t}&=k_3\conc{R}-\prn{k_2+k_{-3}}\conc{I}, \label{equ:I} \\
    \deriv{1}{\conc{P}}{t}&=k_1\conc{R}+k_2\conc{I}, \label{equ:P}
\end{align}
where square brackets denote concentrations, and the infinite sink approximation is applied to P, thereby neglecting reactions originating from P. 
Given boundary conditions $\conc{R}_{t=0}=\conc{R}_0$ and $\conc{I}_{t=0}=0$, the solutions of Eq.~\eqref{equ:R} and \eqref{equ:I} are
\begin{align}
    \conc{R}&=\frac{\conc{R}_0}{\sqrt{A}}\prn{-\rho_-\ex{r_+t}+\rho_+\ex{r_-t}}, \label{equ:R(t)} \\
    \conc{I}&=\frac{\conc{R}_0}{\sqrt{A}}k_3\prn{\ex{r_+t}-\ex{r_-t}}, \label{equ:I(t)}
\end{align}
where $A=\prn{k_1-k_2+k_3-k_{-3}}^2+4k_3k_{-3}$, $r_\pm=\frac{1}{2}\prn{-\prn{k_1+k_2+k_3+k_{-3}}\pm\sqrt{A}}$, and $\rho_\pm=r_\pm+k_1+k_3$. Inserting Eq.~\eqref{equ:R(t)} and \eqref{equ:I(t)} into Eq.~\eqref{equ:P}, we obtain the time evolution of the overall reaction as 
\begin{equation}
    \deriv{1}{\conc{P}}{t}=\frac{\conc{R}_0}{\sqrt{A}}\prn{k_2k_3\prn{\ex{r_+t}-\ex{r_-t}}+k_1\prn{-\rho_-\ex{r_+t}+\rho_+\ex{r_-t}}}.
\end{equation}

The time-dependent overall rate constant is defined by
\begin{equation}
    k\prn{t}\equiv\frac{1}{\conc{R}}\deriv{1}{\conc{P}}{t}.
\end{equation}
Therefore, we find the analytical expression of $k\prn{t}$, which is
\begin{equation}
    k\prn{t}=k_1+k_2k_3\frac{\ex{r_+t}-\ex{r_-t}}{-\rho_-\ex{r_+t}+\rho_+\ex{r_-t}}. \label{equ:kt}
\end{equation}
By taking the limit $t\to+\infty$ to Eq.~\eqref{equ:kt}, the resulting overall rate constant is obtained as
\begin{align}
    k_\mathrm{overall}&\equiv\lim_{t\to+\infty}k\prn{t} \\
    &=k_1-\frac{k_2k_3}{\rho_-} \\
    &=k_1+\frac{k_2}{2k_{-3}}\prn{k_1-k_2+k_3-k_{-3}+\sqrt{\prn{k_1-k_2+k_3-k_{-3}}^2+4k_3k_{-3}}}, \label{equ:k overall}
\end{align}
and the sequential rate constant can be obtained by setting $k_1$ to zero, giving
\begin{equation}
    k_\mathrm{seq}=\frac{k_2}{2k_{-3}}\prn{k_3-k_{-3}-k_2+\sqrt{\prn{k_3-k_{-3}-k_2}^2+4k_3k_{-3}}}.
\end{equation}

\end{appendices}

\bibliography{main}

\newpage
TOC Graphic
\begin{figure}
    \centering
    \includegraphics[width=\linewidth]{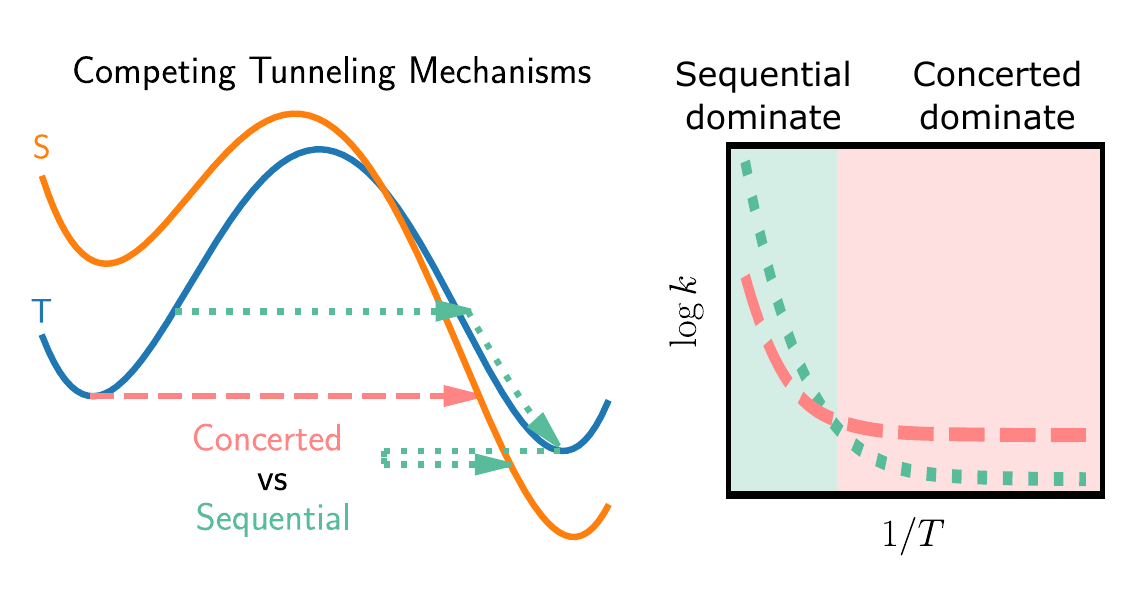}
\end{figure}

\AddToHook{enddocument/afteraux}{%
\immediate\write18{
cp output.aux MAIN.aux
}%
}

\end{document}